\documentclass{iaus}
\usepackage{graphicx}

\title[~~A new model for IRAS F10214+4724] %% give here short title %%
{A new model for the infrared emission of IRAS F10214+4724}

\author[Andreas Efstathiou, Natalie Christopher, Aprajita Verma
 \& Ralf Siebenmorgen]   %% give here short author list %%
{Andreas Efstathiou$^1$, Natalie Christopher$^2$, Aprajita Verma$^3$,
 \and Ralf Siebenmorgen$^4$}

\affiliation{$^1$School of Sciences, European University Cyprus,\\
 Engomi, 1516 Nicosia, Cyprus  \\
 email: {\tt a.efstathiou@euc.ac.cy} \\[\affilskip]
$^2$Oxford Astrophysics, Denys Wilkinson Building, University of Oxford, \\
Keble Rd, Oxford OX1 3RH\\ 
email: {\tt natalie.christopher@astro.ox.ac.uk} \\[\affilskip]
$^3$Oxford Astrophysics, Denys Wilkinson Building, University of Oxford, \\
Keble Rd, Oxford OX1 3RH\\ 
email: {\tt averma@astro.ox.ac.uk} \\[\affilskip]
$^4$European Southern Observatory, Karl-Schwarzschildstr. 2, \\
85748 Garching b. Munchen, Germany\\
email: {\tt rsiebenm@eso.org}
}

\pubyear{2011} %% 
\volume{284}  %% insert here IAU Symposium No.
\pagerange{1--12}
\jname{The Spectral Energy Distribution of Galaxies}
\editors{R.J. Tuffs \&  C.C.Popescu, eds.}
\begin{document}

\maketitle

\begin{abstract}

We present a new model for the infrared emission of the high redshift hyperluminous
infrared galaxy IRAS F10214+4724 which takes into account recent photometric data from
{\em Spitzer} and {\em Herschel} that sample the peak of its spectral energy distribution.
We first demonstrate that the combination of the AGN tapered disc  and starburst models of
Efstathiou and coworkers, while
able to give an excellent fit to the average spectrum of type 2 AGN measured by {\em Spitzer}, fails
to match the spectral energy distribution of IRAS F10214+4724. This is mainly
due to the fact that the $\nu S_\nu$ distribution of the galaxy falls very steeply with
increasing frequency (a characteristic of heavy absorption by dust) but shows
a silicate feature in emission. We propose a model that assumes two components
of emission: clouds that are associated with the narrow-line region and a highly
obscured starburst. The emission from the clouds must suffer significantly stronger
gravitational lensing compared to the emission from the torus to explain the observed
spectral energy distribution.

\end{abstract}

\firstsection % if your document starts with a section,
\section{Introduction}

The title of the discovery paper of IRAS F10214+4724 at a redshift of 2.285
(\cite[Rowan-Robinson et al. 1991]{Rowan-Robinsonetal1991}) was 'An IRAS galaxy with huge luminosity - 
hidden quasar or protogalaxy?'. It soon became clear that the 
reason for the enormous luminosity was that the galaxy was gravitationally
lensed by a factor of 10-100 (e.g. \cite[Broadhurst \& Lehar 1995]{Broadhurst&Lehar95}).
Until recently, however,it was not clear whether the dominant energy source was an active
galactic nucleus (AGN) or a starburst.

 Models for  the infrared spectral energy distribution (SED) of IRAS F10214+4724  
in the mid-90s remained inconclusive (e.g. \cite[Rowan-Robinson et al. 1993]{Rowan-Robinsonetal1993})
mainly due to the limited wavelength coverage. 
Mid-infrared spectrophotometry has been shown for some time now to be       
able to differentiate between a starburst and AGN origin of the infrared      
emission (\cite[Roche et al. 1991]{Roche et al. 1991} and references therein,
 \cite[Genzel et al. 1998]{Genzeletal1998}).   
The mid-infrared spectrum of IRAS F10214+4724 obtained by \cite[Teplitz et al. (2006)]{Teplitzetal2006}  
using the IRS onboard {\em Spitzer} showed a puzzling emission 
feature at rest frame 10$\mu m$ instead of the expected absorption feature  
given the type 2 nature of the AGN suggested by optical observations.  
\cite[Efstathiou (2006)]{Efstathiou2006} proposed that the SED of IRAS F10214+4724 could be    
understood in terms of three components of emission: clouds that are  
associated with the narrow-line region and with a covering factor of  
17\%, an edge-on torus and a highly obscured starburst. It was also noted by \cite[Efstathiou (2006)]{Efstathiou2006},  
however, that because the emission from the torus and the narrow-line  
region clouds arises from diferent regions their magnification may  
be different. 

 In this paper we develop a new model for the infrared emission of  IRAS F10214+4724  
which takes into account new photometric data from {\em Spitzer} and {\em Herschel}   
that cover the wavelength range 70-170$\mu m$ and therefore complete its SED. 

\section{Radiative transfer models}

\subsection{Starburst models}

\cite[Efstathiou, Rowan-Robinson \& Siebenmorgen (2000)]{Efstathiouetal2000} 
developed a starburst model that
has three main features: The first feature of the model
is that it incorporates the stellar population synthesis
model of Bruzual \& Charlot that gives the
spectrum of the stars as a function of their age.
\cite[Efstathiou, Rowan-Robinson \& Siebenmorgen (2000)]{Efstathiouetal2000} 
used the table that assumes a Salpeter IMF and stellar
masses in the range 0.1-125 $M_\odot$. The second important
feature of the model is that it involves detailed
radiative transfer that takes into account multiple
scattering from grains and incorporates a dust
model that includes small transiently heated grains
and PAH molecules as well as large classical grains.
For the calculation of the emission of the small grains
and PAHs the method of \cite[Siebenmorgen \& Kr\"ugel (1992)]{Siebenmorgen&Krugel1992}
was employed. The third feature of the model is
that it incorporates a simple model for the evolution
of the giant molecular clouds that constitute the starburst
once a stellar cluster forms instantaneously at
the center of the cloud. The most important characteristic
of this evolutionary scheme is that by about
10 Myrs after star formation, the expansion of the
HII region leads to the formation of a cold narrow
shell of gas and dust. This naturally explains why the
mid-infrared spectra of starburst galaxies are dominated
by the PAH emission and not by the emission
of hot dust. 
The sequence of molecular clouds at different ages can
be convolved with a star formation history to give the
spectrum of a starburst at different ages.
\cite[Farrah et al. (2003)]{Farrahetal2003} showed that
the starburst models of \cite[Efstathiou, Rowan-Robinson \& 
Siebenmorgen (2000)]{Efstathiouetal2000} in combination with
the tapered discs of \cite[Efstathiou \& Rowan-Robinson (1995)]{Efstathiou&Rowan-Robinson1995},
which are described below, provided good fits to the SEDs
of 41 ultraluminous infrared galaxies.  One
interesting prediction of the \cite[Efstathiou, Rowan-Robinson \& Siebenmorgen (2000)]{Efstathiouetal2000} 
model is that young starbursts show strong near-infrared continuum,
small 6.2$\mu m$ equivalent widths and deep silicate
absorption features and can therefore explain the
galaxies in class 3A of \cite[Spoon et al. (2007)]{Spoonetal2007}. 
The starburst model has recently been revised by 
\cite[Efstathiou \& Siebenmorgen (2009)]{Efstathiou&Siebenmorgen2009}

\subsection{AGN torus models}

The most important constraint on early models for
the torus in AGN (\cite[Pier \& Krolik 1992]{Pier&Krolik1992}, 
\cite[Granato \& Danese 1994]{Granato&Danese1994}, 
\cite[Efstathiou \& Rowan-Robinson 1995]{Efstathiou&Rowan-Robinson1995})
was provided by mid-infrared spectroscopic observations
from the ground in the 8-13$\mu m$ window by
Roche, Aitken and collaborators (\cite[Roche et al. 1991]{Roche et al. 1991}
 and references therein).
The observations showed moderate absorption features
in type 2 AGN and featureless spectra in type 1
AGN. By comparison the flared discs (whose thickness
increases linearly with distance from the central
source) of \cite[Efstathiou \& Rowan-Robinson (1995)]{Efstathiou&Rowan-Robinson1995}) and
 \cite[Granato \& Danese (1994)]{Granato&Danese1994} showed strong
silicate emission features when observed face-on. 
 \cite[Efstathiou \& Rowan-Robinson (1995)]{Efstathiou&Rowan-Robinson1995}) proposed that tapered
discs (whose thickness increases with distance from
the central source in the inner part of the disc but tapers
off to a constant value in the outer disc) with a
density distribution that followed $r^{-1}$ could give flat
spectra in the mid-infrared for an opening angle of
around $45^o$ and an equatorial optical depth at 1000\AA~
of 1000. Models with a smaller opening angle predicted
face-on spectra with shallow silicate absorption
features whereas models with a larger opening angle
showed  emission features. Recent Spitzer observations
showed  emission features in quasars
(e.g. \cite[Hao et al. 2005]{Haoetal2005}, \cite[Hao et al. 2007]{Haoetal2007},
\cite[Siebenmorgen et al 2005]{Siebenmorgenetal2005},
\cite[Spoon et al. 2007]{Spoonetal2007}). The tapered discs of
\cite[Efstathiou \& Rowan-Robinson (1995)]{Efstathiou&Rowan-Robinson1995},
in combination with the starburst
models of \cite[Efstathiou, Rowan-Robinson \& Siebenmorgen (2000)]{Efstathiouetal2000}  have been very
successful in fitting the spectral energy distributions
of a number of AGN even in cases where mid-infrared spectroscopy
was available (\cite[Alexander et al. 1999]{Alexanderetal1999}, 
\cite[Ruiz et al. 2001]{Ruizetal2001}, \cite[Farrah et al. 2003]{Farrahetal2003},
 \cite[Efstathiou \& Siebenmorgen 2005]{Efstathou&Siebenmorgen2005}). 

\begin{figure}[b]
% \vspace*{-2.0 cm}
\begin{center}
 \includegraphics[width=3.4in]{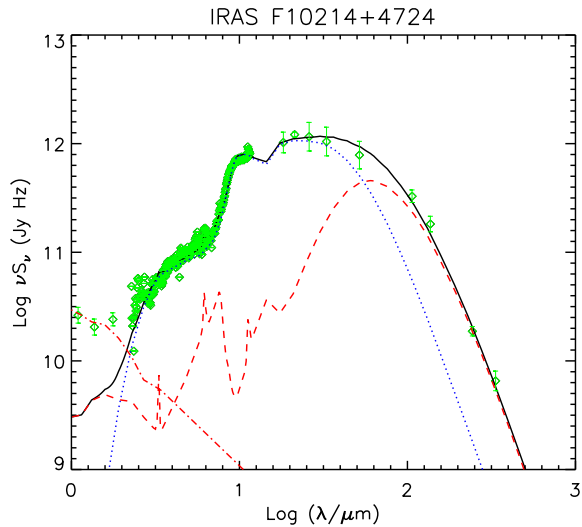} 
% \vspace*{-1.0 cm}
 \caption{
Fit to the spectral energy distribution of IRAS F10214+4724. The dotted and
dashed lines show the contributions of the AGN and starburst respectively
to the total emission (solid line). The dash dotted line also shows a fit
to the IRAC data with a  1Gyr old stellar population. Data from 
\cite[Teplitz et al. (2006)]{Teplitzetal2006}, IRAS,
\cite[Rowan-Robinson et al. (1993)]{Rowan-Robinsonetal1993},
\cite[Sturm et al. (2010)]{Sturmetal2010}.
}
   \label{fig1}
\end{center}
\end{figure}

In this paper we use a grid of tapered disc models
computed with the method of \cite[Efstathiou \& Rowan-Robinson (1995)]{Efstathiou&Rowan-Robinson1995}.
In this grid of models we consider four discrete values
for the equatorial 1000\AA~ optical depth (500, 750, 1000, 1250),
three  values for the ratio of outer to inner disc radii
(20, 60, 100) and three values for the opening angle of the
disc (30$^o$, 45$^o$ and 60$^o$). The spectra are computed for 37 or 74
orientations which are equally spaced in the range 0 to $\pi/2$.   
 
The grid of tapered disc models, as well as the grid of starburst
models discussed above, are available from AE on request
(a.efstathiou@euc.ac.cy). 

\subsection{Comparison with the average spectrum of Seyfert 2s}

 We first explore whether the combination of the tapered disc and
 starburst models described above  can fit the average spectrum
 of Seyfert 2s presented by \cite[Hao et al. 2007]{Haoetal2007}.
 The observed spectrum has been determined by first normalizing
 the spectra at 15$\mu$m and then averaging. 
 The AGN models have also been averaged in an attempt to mimic the
 averaging of the observed spectra. 

 We find that the best fit to the Seyfert 2 average is provided by
a model that assumes a torus opening angle of 45$^o$, an equatorial
1000\AA~ optical depth of 750 and a ratio of outer to inner torus radii
of 20. The best fit starburst model assumes an initial
optical depth of the molecular clouds $\tau_V$ of 200 and a starburst
age of 72Myr.

\section{A new model for IRAS F10214+4724}

We find that the models described above provide a very poor fit to the
SED of  IRAS F10214+4724. This is clearly due to the fact
that the SED of F10214+4724 shows simultaneously characteristics
of almost edge-on and almost face-on emission. 
 The failure of torus emission alone to fit the mid-infrared SED
of F10214+4724 motivated \cite[Efstathiou (2006)]{Efstathiou2006} to propose
a three component model: an edge-on torus, clouds
associated with the narrow-line region and a highly
obscured starburst. The model assumed that the narrow
line region clouds have a covering factor of 17\%
and two discrete values for the dust temperature 
(610 and 200K). The new MIPS and PACS data suggest
that there is about a factor of two more power in
the rest frame wavelength range 20-50$\mu m$ compared with the prediction
of  the model presented in \cite[Efstathiou (2006)]{Efstathiou2006}. Considering
that the new data samples the peak of the SED of    
F10214+4724 this implies that the covering factor
needs to increase even more. It therefore
seems likely that the magnification of the emission
from the narrow-line region is considerably higher
than that from the torus. We find that the new data
can be fitted by assuming an additional component
of narrow-line region dust at a temperature of
80K. The new fit is shown in Figure 1.
The torus emission is assumed to be negligible in this fit.

\end{document}